%% file: main.tex
\documentclass[a4paper]{article}
\usepackage{RR}
\usepackage[french,english]{babel}
\usepackage[latin1]{inputenc}
\usepackage{fontenc}
\usepackage{url} 
\usepackage{hyperref}
\usepackage{amssymb,amsmath, stmaryrd}
\usepackage{algorithmicx}
\usepackage[noend]{algpseudocode}
\usepackage{algorithm}
\usepackage{xspace}
\usepackage{slashbox}
\usepackage{fancybox}
\usepackage{macros}
\usepackage{setspace}

\RRNo{6440}
\RRdate{Janvier  2008}
\RRauthor{
  Pierre Sutra\thanks{LIP6, 104, ave.\ du Président Kennedy, 75016 Paris, France; \url{mailto:pierre.sutra@lip6.fr}} ~~~ Marc Shapiro\\
  {\small Université Paris VI and INRIA Rocquencourt, France}
}
\authorhead{Sutra  \& Shapiro}
\RRtitle{
  Un algoorithme tolérant aux fautes pour la réplication de bases 
  de données dans les systèmes large-échelle
  }
\RRetitle{
  Fault-Tolerant Partial Replication
  in Large-Scale Database Systems
}
\titlehead{}
\RRresume{
  }
\RRabstract{
  \noindent
    We investigate a decentralised approach to committing transactions in a replicated database,
    under partial replication.
    Previous protocols either reexecute transactions entirely
    and/or compute a total order of transactions.
    In contrast, ours applies update values, and    
    generate a partial order between mutually conflicting transactions only.
     It results that
    transactions execute faster,
    and distributed databases commit in small committees.
    Both effects contribute to preserve scalability
    as the number of databases and transactions increase.
    Our algorithm ensures serializability,
    and is live and safe in spite of faults.
}
\RRmotcle
    {
      réplication, large-échelle, base de données, 
    }
\RRkeyword
    {
      data replication, large-scale, database systems
      
    }
\RRprojet{Regal}
\RRtheme{\THCom}
\URRocq 
\begin{document}
\makeRR

\section { Introduction }

Non-trivial consistency problems e.g. 
file systems, collaborative environments, and databases. 
are the major challenge of large-scale systems.
Recently some architectures have emerged to scale file systems up
to thousands of nodes \cite{oceanstore-asplos,muthitacharoen02ivy,DBLP:conf/europar/BuscaPS05}, 
but no practical solution exists for database systems.

At the cluster level protocols based on group communication primitives \cite{1273036,kemme00dont,1107766} 
are the most promising solutions to replicate database systems \cite{1048870} .
In this article we extend the group communication approach to 
large-scale systems.

Highlights of our protocol:

\begin{itemize}

\item Replicas do not re-execute transactions, 
  but apply update values only.

\item We do not compute a total order over of operations.
  Instead transactions are partially ordered.
  Two transactions are ordered only over the data where they conflict.

\item For every transaction $T$ we maintain the graph of $T$'s dependencies.
  $T$ commits locally when $T$ is transitively closed in this graph.

\end{itemize}

The outline of the paper is the following.
Section \ref{sect:model} introduces our model and assumptions.
Section \ref{sect:algorithm} presents our algorithm.
We conclude in Section \ref{sect:end} after a survey of related work.
An appendix follows containing a proof of correctness.

\section { System model and assumptions }
\label{sect:model}

We consider a finite set of asynchronous processes or \emph{sites} \sites,
forming a distributed system.
Sites may fail by crashing, 
and links between sites are asynchronous but reliable.
Each site holds a database that we model as
some finite set of \emph{data items}.
We left unspecified the granularity of a data item.
In the relational model, it can be a column,
a table, or even a whole relational database.
Given a data item $x$, the replicas of $x$, noted \replicas{x},
are the subset of \sites whose databases contain $x$.

We base our algorithm on the three following primitives:%
\footnote{
  Our taxonomy comes from \cite{defago00totally}.
}

\begin{itemize}

\item \emph{Uniform Reliable Multicast} takes as input a unique message $m$
  and a \emph{single} group of sites $g \subseteq \sites$ .
  Uniform reliable multicast consists of the two primitives \rmcast{m} and \rmdeliver{m}.
  With Uniform Reliable Multicast, all sites in $g$ have the following guarantees:
  \begin{itemize}
  \item Uniform Integrity: For every message $m$, every site in $g$
    performs \rmdeliver{m} at most once,
    and only if some site performed \rmcast{m} previously.
  \item Validity: if a correct site in $g$ performs \rmcast{m}
    then it eventually performs \rmdeliver{m}. 
  \item Uniform Agreement: if a site in $g$ performs
    \rmdeliver{m}, then every correct sites in $g$
    eventually performs \rmdeliver{m}.
  \end{itemize}
  Uniform Reliable Multicast is solvable in an asynchronous systems
  with reliable links and crash-prone sites.
  
\item \textit{Uniform Total Order Multicast} takes as input a unique message $m$
  and a single group of sites $g$.
  Uniform Total Order Multicast consists of the two primitives \tomcast{m}
  and \tomdeliver{m}. 
  This communication primitive ensures Uniform Integrity,
  Validity, Uniform Agreement and Uniform Total Order in $g$:
  \begin{itemize}
  \item Uniform Total Order: if a site in $g$ performs \tomdeliver{m}
    and \tomdeliver{m'} in this order,
    then every site in $g$ that performs \tomdeliver{m'} has
    performed previously \tomdeliver{m}.
  \end{itemize}

\item \textit{Eventual Weak Leader Service} Given a group of sites $g$, 
  a site $i \in g$ may call function \wleader{g}.
  \wleader{g} returns a \emph{weak leader} of $g$ :
  \begin{itemize}
  \item $\wleader{g} \in g$.
  \item Let \run be a run of \sites such that a non-empty
    subset $c$ of $g$ is correct in \run.
    It exists a site $i \in c$ and a time $t$ such that for any calls
    of \wleader{g} on $i$ after $t$, \wleader{g} returns $i$.
  \end{itemize}
  This service is strictly weaker than the classical eventual leader service \leader \cite{leaderRaynal},
  since we do not require that every correct site eventually outputs the same leader.
  An algorithm that returns to every process itself, trivially implements
  the Eventual Weak Leader Service.
\end{itemize}

In the following we make two assumptions:
during any run,
\textbf{(A1)} for any data item $x$, at least one replica of $x$ is correct, 
and \textbf{(A2)} Uniform Total Order Multicast is solvable in \replicas{x}.

\subsection { Operations and locks }

\begin{table}[t]\centering\small
  \begin{tabular}{lc}
    & \vspace{0.3cm}\textit{lock held}\\
    \begin{tabular}{l}\textit{lock}\\\textit{requested}\end{tabular} &
    \begin{tabular}{c|c|c|c}
      & $R$ & $W$ & $IW$ \\
      \hline
      $R$  & $1$ & $0$ & $0$  \\
      \hline
      $W$  & $0$ & $0$ & $0$  \\
      \hline
      $IW$ & $0$ & $0$ & $1$  
    \end{tabular}
    \vspace{0.4cm}
  \end{tabular}
  \caption{
    \label{tab:lock}
    Lock conflict table}
\end{table}

Clients of the system (not modeled),
access data items using read and write operations.
Each operation is uniquely identified, and accesses a single data item.
A read operation is a singleton: the data item read,
a write operation is a couple: the data item written, and the update value.%

When an operation accesses a data item on a site, it takes a lock.
We consider the three following types of locks:
read lock (R),
write lock (W),
and intention to write lock (IW).%
%
Table \ref{tab:lock} illustrates how locks conflict with each other;
when an operation requests a lock to access a data item,
if the lock is already taken and cannot be shared,
the request is enqueued in a FIFO queue.
In Table \ref{tab:lock}, $0$ means that the request is enqueued,
and $1$ that the lock is granted.

Given an operation $o$, we note:
\begin{itemize}
\item \ditem{o}, the data item operation $o$ accesses,
\item \isRead{o} (resp. \isWrite{o}) a boolean indicating 
  whether $o$ is a read (resp. a write),
\item and $\replicas{o} \equaldef \replicas{\ditem{o}}$;
\end{itemize}
We say that two operations $o$ and $o'$ \emph{conflict} if
they access the same data item and one of them is a write:
\begin{displaymath}
    \conflict{o}{o'} 
    \equaldef
    \left\lbrace
      \begin{array}{l}
        \ditem{o} = \ditem{o'}\\
        \isWrite{o} \lor \isWrite{o'}
      \end{array}
    \right.
\end{displaymath}

\subsection { Transactions }

Clients group their operations into \emph{transactions}.
A transaction is a uniquely identified set of read and write operations.
Given a transaction $T$,
\begin{itemize}
\item for any operation $o \in T$, function \trans{o} returns $T$,
\item \ro{T} (respectively \wo{T}) is the subset of read (resp. write) operations,
\item \ditem{T} is the set of data items transaction $T$ accesses:
  $\ditem{T} \equaldef \bigunion{o \in T}{\ditem{o}}$.
\item  and $\replicas{T} \equaldef \replicas{\ditem{T}}$.
\end{itemize}

Once a site $i$ grants a lock to a transaction $T$,
$T$ holds it until $i$ commits $T$, $i$ aborts $T$, 
or we explicitly say that this lock is released.

\section { The algorithm }
\label{sect:algorithm}

As replicas execute transactions,
it creates precedence constraints between conflicting transactions.
Serializability theory tell us that this relation must be acyclic \cite{syn:db:1467}.

One solution to this problem,
is given a transaction $T$,
(i) to execute $T$ on every replicas of $T$, 
(ii) to compute the transitive closure of the precedence 
constraints linking $T$ to concurrent conflicting transactions, 
and (iii) if a cycle appears, to abort at least one the transactions
involved in this cycle.

Unfortunately as the number of replicas grows, sites
may crash, and the network may experience congestion.
Consequently to compute (ii) the replicas of $T$ need to agree upon the set
of concurrent transactions accessing \ditem{T}.

Our solution is to use a TO-multicast protocol per data item.

\subsection { Overview }
\label{sect:overview}

To ease our presentation we consider in the following that 
a transaction executes initially on a single site.
Section \ref{sect:variants} generalizes our approach to the case where
a transaction initially executes on more than one site.
We structure our algorithm in five phases:

\begin{itemize}

\item In the  \emph{initial execution phase},
  a transaction $T$
  executes at some site $i$.

\item In the \emph{submission phase},
  $i$ transmits $T$ to \replicas{T}.

\item In the \emph{certification phase}, 
  a site $j$ aborts $T$ if $T$ has read an outdated value.
  If $T$ is not aborted,
  $j$ computes all the precedence constraints
  linking $T$ to transactions previously received at site $j$.

\item In the \emph{closure phase}, $j$ completes its knowledge
  about precedence constraints linking $T$ to others transactions.

\item  Once $T$ is closed at site $j$,
  the \emph{commitment phase} takes place.
  $j$ decides locally whether to commit or abort $T$.
  This decision is deterministic,
  and identical on every site replicating a data item written by $T$.

\end{itemize}

\subsection { Initial execution phase }
\label{sect:init}

A site $i$ executes a transaction $T$ coming from a client
according to the two-phases locking rule \cite{syn:db:1467}, \emph{but}
without applying write operations%
\footnote{
  If $T$ writes a data item $x$ then reads it, we suppose some internals
  to ensure that $T$ sees a consistent value.
}%
.
When site $T$ reaches a commit statement,
it is not committed, instead
$i$ releases $T$'s read locks,
converts $T$'s write locks into intention to write locks,
computes $T$'s update values,
and then proceeds to the submission phase.

\subsection { Submission phase }
\label{sect:com}

In this phase $i$ R-multicasts $T$ to \replicas{T}.
When a site $j$ receives $T$, $j$ marks all $T$'s operations
as pending using variable \pending.
Then if it exists an operation $o \in \pending$,
such that $j=\wleader{\replicas{o}}$,
$j$ TO-multicasts $o$ to \replicas{o}.%
\footnote{
  If instead of this procedure, $i$ TO-multicasts all the operations,
  then the system blocks if $i$ crashes.
  We use a weak leader and a reliable multicast to preserve liveness.
}%

\subsection { Certification phase }
\label{sect:cert}

When a site $i$ TO-delivers an operation $o$ \emph{for the first time}%
\footnote{
  Recall that the leader is eventual, consequently $i$ may receive $o$
  more than one time.
}%
,$i$ removes $o$ from \pending, $i$ certifies $o$.

To certify $o$, $i$ considers any preceding write operations that conflicts with $o$.
We say that a conflicting operation $o'$ \emph{precedes $o$ at site $i$},
\precedesI{o'}{o},
if $i$ TO-delivers $o'$ then $i$ TO-delivers $o$:
\begin{displaymath}
 \precedesI{o'}{o}
 \equaldef
 \left\lbrace
   \begin{array}{l}
     \tomdeliverI{o'} \hb \tomdeliverI{o} \\
     \conflict{o'}{o}
   \end{array}
 \right.
\end{displaymath}
Where given two events $e$ and $e'$,
$e \hb e'$ is the relation $e$ \emph{happens-before} $e'$, 
and \tomdeliverI{o'} is the event: ``site $i$ TO-delivers operation $o'$''.

If $o$ is a read, we check that $o$ did not read an outdated value.
It happens when $o$ executes concurrently to a conflicting write operation $o'$ that is now committed.
Let \committedI be the set of transactions committed at site $i$,
the read operation $o$ aborts,
if it exists an operation $o'$ such that
$\precedesI{o'}{o} \land \trans{o'} \cc \trans{o} \land \trans{o'} \in \committedI$,
where $\trans{o'} \cc \trans{o}$ means that the transactions \trans{o'}  and \trans{o} 
were executed concurrently during the initial execution phase.

If now $o$ is a write, 
$i$ gives an IW lock to $o$: function \forceWriteLock{o}.
If an operation $o'$ holds a conflicting IW lock, 
$o$ and $o'$ share the lock (see Table \ref{tab:lock});
otherwise it means that \trans{o'} is still executing at site $i$,
and function \forceWriteLock{o} aborts it.%
\footnote{
This operation prevents local deadlocks.
}%

\subsection { Precedence graph } 
\label{sect:pgraph}

Our algorithm decides to commit or abort transactions, according to a \emph{precedence graph}.
A precedence graph $G$ is a directed graph where each node is a transaction $T$,
and each directed edge \precedes{T}{T'},
models a precedence constraint
between an operation of $T$, and a \emph{write} operation of $T'$:
\begin{displaymath}
  \precedes{T}{T'}
  \equaldef
  \exists (o,o') \in T \times T',
  \exists i \in \sites,
  \precedesI{o'}{o}
\end{displaymath}
A precedence graph contains also for each vertex $T$
a flag indicating whether $T$ is aborted or not: \isAborted{T}{G},
and the subset of $T$'s operations: \opg{T}{G},
which contribute to the relations linking $T$ to others transactions
in $G$.

Given a precedence graph $G$, 
we note \vertices{G} its vertices set, and \edges{G} its edges set.
Let $G$ and $G'$ be two precedence graphs,
the union between $G$ and $G'$, $G \union G'$, is such that:

\begin{itemize}

\item $\vertices{(G \union G')}=\vertices{G} \union \vertices{G'}$,

\item $\edges{(G \union G')}=\edges{G} \union \edges{G'}$,

\item $\forall T \in \vertices{(G \union G')}$, $\isAborted{T}{(G \union G')}=\isAborted{T}{G} \lor \isAborted{T}{G'}$.

\item $\forall T \in \vertices{(G \union G')}$, $\opg{T}{(G \union G')}=\opg{T}{G} \union \opg{T}{G'}$.

\end{itemize}

We say that $G$ is a subset of $G'$, noted $G \subseteq G'$, if:

\begin{itemize}

\item $\vertices{G} \subseteq \vertices{G'} \land \edges{G} \subseteq \edges{G'}$,

\item $\forall T \in \vertices{G}$, $\isAborted{T}{G} \implies \isAborted{T}{G'}$,

\item $\forall T \in \vertices{G}$, $\opg{T}{G} \subseteq \opg{T}{G'}$.

\end{itemize}

Let $G$ be a precedence graph, \incomingNeighbors{T}{G} 
(respectively \outgoingNeighbors{T}{G}) is the restriction
of \vertices{G} to the subset of vertices formed
by $T$ and its incoming (resp. outgoing) neighbors.
The \emph{predecessors} of $T$ in $G$: \predecessors{T}{G},
is the precedence graph representing the transitive closure of the dual of
the relation $\edges{G}$ on $\{T\}$.

\subsection { Deciding }

\begin{algorithm}[t]\small
  \caption{\label{alg:decide}
    \decide{T}{G}, code for site $i$}
  \begin{algorithmic}[1]
    
    \Variable $G' \assign \emptyGraph$ \Comment{ a directed graph }
    \State

    \ForAll { $C \subseteq \cycles{G}$ }
      \If { $\forall T \in C, \neg \isAborted{T}{G}$ }
        \State $G' \assign G' \union C$
      \EndIf
    \EndFor
    
    \If { $T \in \breakCycles{G'}$ } \label{alg:decide:cb}
      \Return \false
    \Else
      \Return \true
    \EndIf

  \end{algorithmic}
\end{algorithm}

Each site $i$ stores its own precedence graph \Gi,
and decides locally to commit or abort a transaction according to it.
More precisely $i$ decides according to the graph \predecessorsI{T}.
For any cycle $C$ in the set of cycles in \predecessorsI{T}: \cycles{\predecessorsI{T}}.
$i$ must abort at least one transaction in $C$.
This decision is deterministic,
and $i$ tries to minimize the number of transactions aborted.

Formally speaking $i$ solves the minimum feedback vertex set problem
over the union of all cycles in \predecessorsI{T}
containing only non-aborted transactions
The minimum feedback vertex set problem is an NP-complete optimization problem, and 
the literature about this problem is vast \cite{574848}.
We consequently postulate the existence of an heuristic: \breakCycles{}.
\breakCycles{} takes as input a directed graph $G$,
and returns a vertex set $S$ such that $G \setminus S$ is acyclic.

Now considering a transaction $T \in \Gi$ such that $G=\predecessorsI{T}$,
Algorithm~\ref{alg:decide}
returns \false if $i$ aborts $T$, or \true otherwise.

\subsection{ Closure phase }

In our model sites replicate data partially,
and consequently maintain an incomplete view of the
precedence constraints linking transactions in the system.
Consequently they need to complete their view
by exchanging parts of their graphs.
This is our closure phase:

\begin{itemize}

\item When $i$ TO-delivers an operation $o \in T$,
  $i$ adds $T$ to its precedence graph, and adds $o$ to \opg{T}{\Gi}.
  Then $i$ sends $\predecessorsI{T}$ to
  \replicas{\outgoingNeighborsI{T}} (line \ref{alg:lsdb:7}).

\item When $i$ receives a precedence graph $G$,
  if $G \not\subseteq \Gi$,
  for every transaction $T $ in \Gi, such that $\predecessors{T}{G} \not\subseteq \predecessorsI{T}$,
  $i$ sends $\predecessors{T}{G \union \Gi}$ to \replicas{\outgoingNeighborsI{T}}.
  Then $i$ merges $G$ to \Gi (lines \ref{alg:lsdb:8} to \ref{alg:lsdb:10}).

\end{itemize}

Once $i$ knows all the precedence constraints linking
$T$ to others transactions, 
we say that $T$ is \emph{closed} at site $i$.
Formally $T$ is closed at site $i$ 
when the following fixed-point equation is true at site $i$:
\begin{displaymath}
  \closedI{T}
  =
  \left\lbrace
  \begin{array}{l}
    \opg{T}{G} = T \\
    \forall T' \in \vertices{\incomingNeighborsI{T}}, \closedI{T'}
  \end{array}
  \right.
\end{displaymath}

Our closure phase ensures that during every run \run,
for every correct site $i$, and every transaction $T$
which is eventually in \Gi,
$T$ is eventually closed at site $i$.

\begin{algorithm}\small
  \caption{\label{alg:lsdb}
   code for site $i$}
  \begin{algorithmic}[1]

    \Variables  $ \Gi \assign \emptyGraph$; $\pending \assign \emptyset $
    \State 

    \Loop \Comment{Initial execution}
      \State let $T$ be a new transaction 
      \State \initialExecution{T}
      \If { $  \wo{T} \neq \emptyset  $  }
        \State \rmcast{T} to \replicas{T}
      \Else
        \State \commit{T} \label{alg:lsdb:0}
      \EndIf
    \EndLoop
    \State 

    \When { \rmdeliver{T} } \label{alg:lsdb:1} \Comment{Submission}
      \ForAll { $ o \in T : i \in \replicas{o} $ }
        \State $ \pending \assign \pending \union \{ o \} $
      \EndFor \label{alg:lsdb:2}
    \EndWhen

    \When { $ \exists o \in \pending \land  i = \wleader{\replicas{o}} $}  \label{alg:lsdb:3}
        \State \tomcast{o} to \replicas{o}  \label{alg:lsdb:4}
    \EndWhen

    \WhenFT { $ \tomdeliver{o} $ } \label{alg:lsdb:4b} \Comment{Certification}
      \State $ \pending \assign \pending \setminus \{ o \} $
      \State let $ T=\trans{o} $
      \State $ \vertices{\Gi} \assign \vertices{\Gi} \union \{  T \} $
      \State $ \opg{T}{\Gi} \assign \opg{T}{\Gi} \union \{o\} $
      \If { $ \isRead{o} \land (\exists o', \precedesI{o'}{o} \land \trans{o'} \cc \trans{o} \land \trans{o'} \in \committedI)$ }
        \State \setAbortedI{T} \label{alg:lsdb:4c}
      \ElsIf { $ \isWrite{o} $ }
        \State \forceWriteLock{o}    
        \ForAll{ $ o': \precedesI{o'}{o} $ } \label{alg:lsdb:5}
            \State $ \edges{\Gi} \assign \edges{\Gi} \union \{ (\trans{o'}, T) \} $ \label{alg:lsdb:6}
        \EndFor
      \EndIf
      \State \send{\predecessorsI{T}} to \replicas{\outgoingNeighborsI{T}} \label{alg:lsdb:7}
    \EndWhenFT

    \When { \receive{T,G} } \label{alg:lsdb:8} \Comment{Closure}
    \ForAll { $T \in \Gi$ }
        \If { $\predecessors{T}{G} \not\subseteq \predecessorsI{T} $ }
            \State \send{\predecessors{T}{\Gi \union G}} to \replicas{\outgoingNeighborsI{T}} \label{alg:lsdb:9}
        \EndIf
      \EndFor
    \State $\Gi \assign \Gi \union G$ \label{alg:lsdb:10}
    \EndWhen

    \When {
      $
          \exists T \in \Gi,
          \left\lbrace
          \begin{array}{l}
            i \in \replicas{\wo{T}} \\
            \closedI{T} \\
            \holdIwLocks{T}
          \end{array}
          \right.
      $ 
    } \Comment{Commitment} \label{alg:lsdb:11}
      \If { $ \neg~\isAbortedI{T} \land \decide{T}{\predecessorsI{T}} $ }
        \State \commit{T}
      \Else
        \State \abort{T}
      \EndIf
    \EndWhen 

   \end{algorithmic}
\end{algorithm}

\subsection { Commitment phase }
\label{sect:commit}

If $T$ is a read-only transaction: $\wo{T}=\emptyset$,
$i$ commits $T$ as soon as $T$ is executed (line \ref{alg:lsdb:0}).

If $T$ is an update, $i$ waits that $T$ is closed and
holds all its IW locks: function \holdIwLocks{} 
(line \ref{alg:lsdb:10}).
Once these two conditions hold, $i$ computes \decide{T}{\predecessorsI{T}}.
If this call returns \true, $i$ commits $T$:
for each write operation $o \in \wo{T}$,
with $i \in\replicas{o}$,
$i$ considers any write operation $o'$ such that
$\precedes{T}{\trans{o'}} \in \Gi \land \conflict{o}{o'}$.
If \trans{o'} is already committed at site $i$, $i$ does nothing;
otherwise $i$ applies $o$ to its database.

Algorithm~\ref{alg:lsdb} describes our algorithm.
This protocol provides serializability
for partially replicated database systems:
any run of this protocol is equivalent to a run 
on a single site \cite{syn:db:1467}.
The proof of correctness appears 
in Appendix.

\subsection { Initial execution on more than one site } 
\label{sect:variants}

When initial execution phase does not take place on a single site 
we compute the read-from dependencies.
More precisely when a site $i$ receives a read $o$ accessing a data item 
it does not replicate,
$i$ sends $o$ to some replica $j \in \replicas{o}$.
Upon reception $j$ executes $o$.
At the end of execution $j$ sends back to $i$ the transitive closure 
containing read-from dependencies and starting from $T$.

Once $i$ has executed locally or remotely all the read operations in $T$,
$i$ checks if the resulting graph contains cycles in which $T$
is involved.
If this is the case, $T$ will be aborted, and instead of submitting it,
$i$ re-executes at least one of $T$'s read operations
Otherwise $i$ computes the write set and the update values,
and sends $T$ with its read-from dependencies by Uniform Reliable Multicast.
The dependencies are merged to precedence graph when a site receives
an operation by Total Order Multicast.
The rest of the algorithm remains the same.

\subsection { Performance analysis }

We consider Paxos \cite{Lam06} as a solution to Uniform Total Order Multicast.
Since precedence constraints in a cycle are \emph{not} causally related,
\lsdb achieves a message delay of $5$:
$2$ for Uniform Reliable Multicast, and $3$ for Uniform Total Order Multicast.
It reduces to $4$, if in each replica group the leader of Paxos is also the weak leader of $g$.

Let $o$ be the number of operations per transaction,
and $d$ be the replication degree,
the message complexity of \lsdb is $5od+(od)^2$: 
$2od$ for Uniform Reliable Multicast,
$o$ Uniform Total Order Multicasts, each costing $2d$ messages,
and $od$ replicas execute line \ref{alg:lsdb:7}, each site sending $od$ messages.
Again, if in each replica group, the leader of Paxos is also the weak leader of $g$,
the message complexity of our protocol reduces to $4od+(od)^2$

\section { Concluding remarks } 
\label{sect:end}

\subsection { Related work }

Gray et al. \cite{233330} prove
that scale traditional eager and lazy replications does not scale:
the deadlock rate increase as the cube of the number of sites, 
and the reconciliation rate increases as the square.
Wiesmann and Schiper confirm practically this result \cite{1048870}.
Fritzke et al. \cite{879303} propose a replication scheme 
where sites TO-multicast each operations and execute them upon reception.
However they do not prevent global deadlocks with a priority rule;
it increases abort rate.
Preventive replication \cite{1107766} considers that a bound 
on processor speed, and network delay is known.
Such assumptions do not hold in a large-scale system.
The epidemic algorithm of Holiday et al \cite{holliday02partial}
aborts concurrent conflicting transactions
and their protocol is not live in spite of one fault.
In all of these replication schemes, 
each replica execute all the operations accessing 
the data items it replicates.
Alonso proves analytically that it reduces the
scale-up of the system \cite{alonso97partial}.

The DataBase State Machine approach \cite{citeulike:552795}
applies update values only but in a fully replicated environment.
Its extensions \cite{LABOS-CONF-2006-024,sousa01partial} 
to partial replication require a total order over transactions.

Committing transactions using a distributed serialization graph
is a well-known technique \cite{shih90survey}.
Recently Haller et al. have proposed to apply it \cite{1099563}
to large-scale systems, but their solution does not handle replication, nor faults.

\subsection { Conclusion } 

We present an algorithm for replicating database systems in a large-scale system.
Our solution is live and safe in presence of non-byzantine faults.
Our key idea is to order conflicting transaction per data item,
then to break cycles between transactions.
Compared to previous existing solutions, ours either achieves lower latency
and message cost, or does not unnecessarily abort concurrent conflicting transactions.

The closure of constraints graphs is a classical idea in distributed systems.
We may find it in the very first algorithm about State Machine
Replication \cite{359563}, or in a well-known algorithm to solve
Total Order Multicast \cite{defago00totally}.%
\footnote{
  In \cite{359563} Lamport closes the $\ll$ relation for every request
  to the critical section.
  In \cite{defago00totally} the total order multicast protocol attributed to Skeen,
  closes the order over natural numbers to TO-multicast a message.
}%
We believe that the closure generalizes to a wider context,
where a constraint is a temporal logic formula over sequences of 
concurrent operations.

\label{lastpage}

\bibliographystyle{plain}
{\footnotesize
\renewcommand{\url}[1]{}
\bibliography{bib,main}
}

\onecolumn
\newpage
\appendix
\input{proof}
\end{document}

%% file: proof.tex
\subsection { Additionnal notations }

We note \data the universal set of data item, 
\transactions the universal set of transactions, 
and \precedenceGraphs the universal set of precedence graphs
constructed upon \data.

Let \run be a run of Algorithm~\ref{alg:lsdb},
given a site $i$ we note $\textit{event}_{i}$ when the event \textit{event}
happens at site $i$ during \run;
moreover if \textit{value} is the result of this event we note it:
$\textit{event}_i=\textit{value}$.

Let \run be a run of Algorithm~\ref{alg:lsdb}, we note:

\begin{itemize}

\item \faulty{\run} the set of sites that crashes during \run, 

\item \correct{\run} the set $\sites \setminus \faulty{\run}$.

\item \committed{\run} the transactions committed during \run, i.e. $\{ T \in \transactions, \exists i \in \sites, T \in \committedI \}$, 

\item and \aborted{\run} the transactions aborted during \run, i.e. $\{ T \in \transactions, \exists i \in \sites, T \in \abortedI \}$.

\end{itemize}

Given a site $i$ and a time $t$,
we note \Git the value of \Gi at time $t$.

\subsection { Proof of correctness  }

Since the serializability theory is over a finite set of transactions,
we suppose hereafter that during \run a finite subset of \transactions is sent to the system.

Let \run be a run of Algorithm~\ref{alg:lsdb}, we now proove a series of propositions
leading to the fact that \run is serializable.

\bigskip

\begin{proposition}
  \label{prop:1}
  \begin{multline*}
  \forall T \in \transactions,
  (
  \exists j \in \sites,
  \rmdeliverSite{T}{j} \in \run
  )\\
  \implies
  (
  \forall o \in T, \forall i \in \replicas{o} \inter \correct{\run},\tomdeliverSite{o}{i} \in \run
  )
\end{multline*}
\end{proposition}

\begin{preuve}

Let $T$ be a transaction and $j$ a site that R-delivers $T$ during \run.

\begin{list}{}{}

\item \textbf{F1.1} $ \forall i \in \replicas{T} \inter \correct{\run}, \rmdeliverSite{T}{i} $
  \begin{list}{}{}
  \item By the Uniform Agreement property of Uniform Reliable Multicast.
  \end{list}

\item \textbf{F1.2} $ \forall o \in T, \exists k \in \correct{\run} \inter \replicas{o}, \tomcastSite{o}{k} \in \run $
  \begin{list}{}{}
  \item \textbf{F1.2.1}
    $
    \exists l \in \correct{\run} \inter \replicas{o}, \wleaderSite{\replicas{o}}{l}=l \land \rmdeliverSite{o}{l}
    $
    \begin{list}{}{}
    \item By fact \textbf{F.1.1}, assumption \textbf{A1} and the properties of the Eventual Weak Leader Service.
    \end{list}
  \item By fact \textbf{F1.2.1} eventually a correct site executes line \ref{alg:lsdb:4} in Algorithm~\ref{alg:lsdb}.
  \end{list}
  
\end{list}

Fact \textbf{F1.2} and the Validity and the Agreement properties of Total Order Multicast conclude our claim.

\end{preuve}

In the following we say that a transaction $T$ is submitted to the system:
$T \in \submitted{\run}$, if a site i R-delivers $T$ during \run.\\

\begin{proposition}
  \label{prop:2}
  \begin{multline*}
    \forall T \in \submitted{\run}, \forall i \in \replicas{T}, \forall o \in T, \\
    \exists G \in \precedenceGraphs, o \in \opg{T}{G} \land \receiveSite{G}{i}
  \end{multline*}
\end{proposition}

\begin{preuve}

  \begin{list}{}{}
    \item \textbf{F2.1} $\forall i \in \sites, \forall t,t', t > t' \implies \Git \subseteq \pgraphSiteTime{i}{t'}$
    \item \textbf{F2.2} $\forall G \in \precedenceGraphs, \forall T \in \transactions, T \in G \implies T \in \predecessors{T}{G}$
      \begin{list}{}{}
      \item By definition of \predecessors{T}{G}.
      \end{list}
  \end{list}

  By proposition \textbf{P1}, facts \textbf{F2.1} and \textbf{F2.2}, and since links are reliable.

\end{preuve}

\begin{proposition}
  \label{prop:3}
    \begin{multline*}
    \forall T, T' \in \submitted{\run},\\
    \precedes{T}{T'}
    \implies
    (
    \exists o,o' \in T \times T', \exists i \in \correct{\run},
    \precedesSite{o}{o'}{i}
    )
    \end{multline*}
\end{proposition}

\begin{preuve}

By definition of \precedes{T}{T'}, let $o,o' \in T \times T'$ and 
let $j$ be a site such that \precedesSite{o}{o'}{j}.
Since \conflict{o}{o'} and an operation applies on a single data item,
we note $x$ the unique data item such that $x=\ditem{o}=\ditem{o'}$.

\begin{list}{}{}

\item \textbf{F3.1} $j \in \replicas{x}$
  \begin{list}{}{}
  \item Site $j$ TO-delivers $o$ during \run and links are reliable.
  \end{list}

\item \textbf{F3.2} $\exists i \in \replicas{x} \inter \correct{\run}, \tomdeliverSite{o}{i} \land \tomdeliverSite{o'}{i}$
  \begin{list}{}{}
  \item By assumption \textbf{A1} $\exists i \in \replicas{x} \inter \correct{\run}$, 
    and by the Uniform Agreement property of Total Order Multicast, since $i$ is correct during \run,
    $i$ TO-delivers both $o$ and $o'$.
  \end{list}

\end{list}

Fact \textbf{F3.2} and the Total Order property of Total Order Multicast concludes our claim.

\end{preuve}

\begin{proposition}
  \label{prop:4}
    \begin{multline*}
    \forall T \in \submitted{\run}, \forall i \in \sites\\
    (
    \exists t, T \in \Git
    )
    \implies
    (
    \exists T_1, \ldots, T_{m \geq 0} \in \submitted{\run},
    i \in \replicas{T_m} \land T \precedesRelation T_1 \precedesRelation \ldots \precedesRelation T_m
    )
    \end{multline*}
\end{proposition}

\begin{preuve}

  Since $\pgraphSiteTime{i}{0}=(\emptyset,\emptyset)$, let us consider the first time $t_0$ at which 
  $T \in \Git$.
  
  According to \lsdb either:

  \begin{itemize}

    \item $i$ TO-delivers an operation $o \in T$ at $t_0$, and thus $i \in \replicas{T}$. \qed

    \item or $i$ receives a precedence graph $G'$ from a site $j$ such that $T \in G'$.
      Now since links are reliable, note $t_1$ the time at which $j$ send $G'$ to $i$.
      According to lines \ref{alg:lsdb:7} and \ref{alg:lsdb:9}, it exists a transactions $T'$
      such that $T \in \predecessors{T'}{\pgraphSiteTime{j}{t_1}}$, and a transaction $T''$
      such that $T'' \in \outgoingNeighbors{T'}{\pgraphSiteTime{j}{t_1}}$ and $i \in \replicas{T''}$.
      
      From $T \in \predecessors{T'}$, by definition of the predecessors,
      we obtain $T \precedesRelation \ldots \precedesRelation T'$,
      and from $T'' \in \outgoingNeighbors{T'}{\pgraphSiteTime{j}{t_1}}$ we obtain \precedes{T'}{T''}.
      Thus $T \precedesRelation \ldots \precedesRelation T' \precedesRelation T''$, 
      with $i \in \replicas{T''}$.

  \end{itemize}

\end{preuve}

\begin{proposition}
  \label{prop:5}
  \begin{multline*}
    \forall T \in \submitted{\run}, \forall i \in \correct{\run},\\
    (
    \exists t, T \in \Git
    )
    \implies
    (
    \exists t, \opg{T}{\Git} = T
    )
  \end{multline*}
\end{proposition}

\begin{preuve}

  Let $T_0$ be a transaction submitted during \run and let $i$ be a site that eventually 
  hold $T_0$ in \Gi.

  By proposition \textbf{P4} it exists $T_1, \ldots, T_{m \geq 0} \in \submitted{\run}$ such
  that $i \in replicas{T_m}$ and $T \precedesRelation T_1 \precedesRelation \ldots \precedesRelation T_m$.

  Let $k \in \llbracket 0,m \rrbracket$, we note $\mathcal{P}(k)$ the following
  property:
  
  \begin{displaymath}
    \mathcal{P}(k) \equaldef \forall j \in \correct{\run} \inter \replicas{T_k}, \exists t \in \opg{T_0}{\pgraphSiteTime{j}{t_0}} = T_0
  \end{displaymath}

  Observe that by proposition \textbf{P2} $\mathcal{P}(0)$ is true.
  We now proove that $\mathcal{P}(k)$ is true for all the $k$ by induction:

  Let $o,o' \in T_k \times T_{k+1}$, and $j \in \correct{\run}$ such that
  \precedesSite{o}{o'}{j}.
  
  Let $t_0$ be the first time at which $j$ TO-delivers $o$ during \run.
  
  Let $t_2$ be the first time at which $\opg{T_k}{\pgraphSiteTime{j}{t}}=T_k$ 
  (since $\pgraphSiteTime{j}{0}=(\emptyset,\emptyset)$, and $\mathcal{P}(k)$ is true).
  
  Let $t_1$ be the first time at which $j$ To-delivers $o'$ during \run.
  
  Observe that since \precedesSite{o}{o'}{j}, $t_O < t_1$.
  It follow that we have three cases to consider:
  
  \begin{itemize}
    
  \item cases $t_2 < t_0 < t_1$ and $ t_0 < t_2 < t_1$\\
    
    In these cases when $j$ To-delivers $o'$, we have:
    
    \begin{displaymath}
      \precedes{T_k}{T_{k+1}} \in \pgraphSiteTime{j}{t_1} 
      \land
      \opg{T_k}{\pgraphSiteTime{j}{t_1} = T_k}
    \end{displaymath}

    Thus,

    \begin{displaymath}
      T_k \in \predecessors{T_{k+1}}{\pgraphSiteTime{j}{t_1}} 
      \land
      \opg{T_k}{\predecessors{T_k}{\pgraphSiteTime{j}{t_1}}} = T_k
    \end{displaymath}
    
    and according to \lsdb, $j$ sends \predecessors{T_{k+1}}{\pgraphSiteTime{j}{t_1}}
    to \replicas{\outgoingNeighbors{T_{k+1}}}{\pgraphSiteTime{j}{t_1}}.

    Now since 
    $\replicas{T_{k+1}} \subseteq \replicas{\outgoingNeighbors{T_{k+1}}}{\pgraphSiteTime{j}{t_1}}$,
    given a site $j \in \replicas{T_{k+1}}$, eventually $j$ receives 
    \predecessors{T_{k+1}}{\pgraphSiteTime{j}{t_1}}, and merges it into its own precedence graph.\\
    
  \item case $t_0 < t_1 < t_2 $\\

    We consider two-subcases:

    \begin{itemize}
      
    \item At $t_2$ $j$ delivers an operation of $T_k$, and this operation is different from
      $o'$. Now since $\precedes{T_{k}}{T_{k+1}} \in \pgraphSiteTime{j}{t_2}$, $\mathcal{P}(k+1)$ is true.
      
    \item If now $j$ receives a graph $G$ such that $\opg{T_k}{G} = T_k$, by definition of $t_2$,
      $G \subseteq \pgraphSiteTime{j}{t_2}$, and more precisely, 
      $\predecessors{T_k}{G} \not\subseteq \predecessors{T_k}{\pgraphSiteTime{j}{t_2}}$.

      It follows that $j$ sends $\predecessors{T_k}{G \union \pgraphSiteTime{j}{t_2}}$ to 
      $\replicas{\outgoingNeighbors{T_k}{G \union \union \pgraphSiteTime{j}{t_2}}}$.
      Finally since by definition of $t_1$, $\precedes{T_k}{ T_{k+1}} \in \pgraphSiteTime{j}{t_2}$,
      we obtain $T_{k+1} \in \outgoingNeighbors{T_{k}}{G \union \pgraphSiteTime{j}{t_2}}$,
      from which we conclude that  $\mathcal{P}(k+1)$ is true.

    \end{itemize}
    
  \end{itemize}

  To conclude observe that since $i \in \replicas{T_m}$ and $\mathcal{P}(m)$ is true,
  eventually $\opg{T_0}{\pgraphSiteTime{i}{t_0}} = T_0$.

\end{preuve}

\begin{proposition}
  \label{prop:6}
      \begin{multline*}
    \forall T \in \submitted{\run}, \forall i \in \correct{\run},\\
    (
    \exists t, T \in \Git
    )
    \implies
    (
    \exists t, \forall T' \in \submitted{\run},
    \precedes{T'}{T} \implies (T',T) \in \Git
    )
  \end{multline*}
\end{proposition}

\begin{preuve}

\begin{list}{}{}

\item \textbf{F6.1}
  $
  \forall T,T' \in submitted{\run}, \forall o,o' \in T \times T',
  (\exists i \in \sites, \precedesSite{o'}{o} \implies \forall j \in \replicas{o}, \precedesSite{o}{o'}{j}
  $
  \begin{list}{}{}
  \item By the Uniform Agreement and Total Order properties of Total Order MBroadcast
  \end{list}

\item \textbf{F6.2}
  $
  \forall T \in \submitted{\run}, \forall o \in T, \forall i \in \correct{\run},
  (
  \exists t, o \opg{T}{\Git}
  ) 
  \implies
  (
  \forall T' \submitted{\run},
  \precedes{T'}{T}
  \implies 
  \exists t, (T,T') \in \Git
  )
  $
  
  \begin{list}{}{}
    
  \item Since $o \in \opg{T}{\Git}$ and $\pgraphSiteTime{i}{0}=(\emptyset,\emptyset)$, either:

  \begin{enumerate}

  \item $i \in \replicas{T} \land \tomdeliverSite{i}{o}$

    First observe that since links are reliable $i \in \replicas{o}$.

    Let $T'$ be a a transaction, $o' \in T'$ an operation, and $k$ a site
    such that \precedesSite{o'}{o}{k}.

    By fact \textbf{F6.1} since $i,j \in \replicas{o}$, \precedesSite{o'}{o}{i}.\\
    
  \item $\exists G \in \precedenceGraphs, \receiveSite{G}{T} \land o \in \opg{T}{G}$

    According to \lsdb it exists $k_0,\ldots,k_m$ sites
    sucht that:
    
    \begin{itemize}

    \item $k_0$ TO-delivers $o$ during \run, and execute line \ref{alg:lsdb:7} 
      sending \predecessors{T}{\pgraphSite{k_0}} with $o \in \opg{T}{predecessors{T}{\pgraphSite{k_0}}}$
      and $k_1 \in \replicas{\outgoingNeighbors{T}{\pgraphSite{k_0}}}$.

    \item $k_1$ receives \predecessors{T}{\pgraphSite{k_0}} during \run and then
      execute line \ref{alg:lsdb:7} or line \ref{alg:lsdb:9}, sending a precedence graph
      $G$ such that $\predecessors{T}{\pgraphSite{k_0}}\subseteq G$ to a set of replicas containinig $k_2$.

    \item etc ... until $i$ receives it.

    \end{itemize}

    Consequently $\predecessors{T}{\pgraphSite{k_0}} \subseteq \Git$, and according to our reasonning in item 
    1, we conclude that fact \textbf{F6.2} is true.
    
  \end{enumerate}
    
  \end{list}

\end{list}

Fact \textbf{F6.2} and proposition \textbf{P5} conclude.  
  
\end{preuve}

We are now able to proove our central theorem: every transaction is eventually closed
at a correct site.\\

\begin{theorem}
  \label{theo:1}
  \begin{multline*}
    \forall T \in \submitted{\run}, \forall i \in \correct{\run},\\
    (
    \exists t, T \in \Git
    )
    \implies
    (
    \exists t, \closed{T}{\Git}
    )
      \end{multline*}
\end{theorem}

\begin{preuve}

We consider that a finite subset of \transactions are sent to the system,
consequently \submitted{\run} is also finite.
Let $C_{T}$ be the graph resulting from
the transitive closure of the relation \precedesRelation on $\{T\}$.
According to proposition \textbf{P6}, $C_{T}$ is eventually in 
\Git, and thus according to proposition \textbf{P5},
$T$ is eventually closed at site $i$.  

\end{preuve}

\begin{proposition}
  \label{prop:7}
  \begin{multline*}
  \forall T \in \submitted{\run}, \forall i, j \in \sites, \forall t, t',\\
  (
  T \in \Git \land T \in \pgraphSiteTime{j}{t'} \land \closed{T}{\Git} \land \closed{T}{\pgraphSiteTime{j}{t'}}
  ) 
  \implies 
  (
  \predecessors{T}{\Git} = \predecessors{T}{\pgraphSiteTime{j}{t'}}
  )
  \end{multline*}
\end{proposition}

\begin{preuve}

\begin{list}{}{}
  
  \item \textbf{F7.1} $ \vertices{\predecessors{T}{\Gi}} = \vertices{\predecessors{T}{\pgraphSite{j}}} $
    \begin{list}{}{}
    \item 
      Let $T' \in \predecessors{T}{\Gi}$.
      By definition it exists $T_1, \ldots, T_m$ such that 
      $T' \precedesRelation T_1 \precedesRelation \ldots \precedesRelation T_m \precedesRelation T \subseteq \Gi$.
      By an obious induction on $m$ using proposition \textbf{P6} we conclude that $T'$ is also
      in $\predecessors{T}{\pgraphSite{j}}$.
    \end{list}

  \item \textbf{F7.2} $ \edges{\predecessors{T}{\Gi}} = \edges{\predecessors{T}{\pgraphSite{j}}} $
    \begin{list}{}{}
    \item 
      Identical to the reasonning proposed for fact \textbf{F7.1}.
    \end{list}

  \item \textbf{F7.2} $ \forall T' \in \predecessors{T}{\Gi}, \opg{T'}{\predecessors{T}{\Gi}} = \opg{T'}{\predecessors{T}{\pgraphSite{j}}} $
    \begin{list}{}{}
    \item 
      By fact \textbf{F7.1} and since $T$ is closed at both sites $i$ and $j$.
    \end{list}

  \item \textbf{F7.4}
    $ 
    \{ T' | \isAborted{T}{\Gi} \}
    =
    \{ T' | \isAborted{T}{pgraphSite{j}} \} 
    $
    \begin{list}{}{}
    \item 
      Let $T' \in \predecessors{T}{\Gi}$ such that $\isAborted{T'}{\predecessors{T}{\Gi}}$.
      According to \lsdb, it exists a site $k$ and a read operation $r \in T'$ such that 
      $k$ TO-delivers $r$ during \run, and then $k$ set the aborted flag of $T'$ in its
      precedence graph.

      Now let $k'$ be a replica of $r$, by the Uniform Agreement and the Total Order Property of
      Total Order Multicast, when $k'$ TO-delivers $r$, it also set the aborted flag of $T'$
      in its precedence graph.
    \end{list}     
\end{list}
  
By the conjunction of facts \textbf{F7.1} to \textbf{F7.4}.

\end{preuve}

We proove now that \run is serializable \cite{syn:db:1467}.

Let $O(x,\run)$ be the set of write operation over the data item $x$ during \run,
we define the relation \versionOrder as follows:
\begin{displaymath}
  \forall x \in \data, \forall o_1,o_2 \in O(x,\run),\\
  x_1 \versionOrder x_2 \equaldef \exists i \in \replicas{x}, \precedesSite{o}{o'}{i}
\end{displaymath}

\begin{proposition}
  \label{prop:8}
  $
  \versionOrder \text{ is a version order for } \run.
  $
  \bigskip

\end{proposition}

\begin{preuve}

  Let $O(x,\run)$ be the set of write operation over the data item $x$ during \run;
  and let $i \in \replicas{x} \inter \correct{\run}$ (assumption \textbf{A1}).

  According to \lsdb $o$ is executed only if \trans{o} is committed during \run
  consequently $i$ commits during \run any transaction $T$ such that 
  $\exists o \in \wo{T}, O(x,\run)$.
  Consequently \versionOrder is total over $O(x,\run)$, and by the 
  Total Order and Uniform Agreement properties of Total Order Multicast,
  \versionOrder is an order over $O(x,\run)$.

\end{preuve}

\begin{proposition}
  \label{prop:9}
  \begin{multline*}
    \forall T, T' \in \mvsg{\run}{\versionOrder}, \\
    (
    (T,T') \in \mvsg{\run}{\versionOrder}
    \land \wo{T} \neq \emptyset
    \land \wo{T'} \neq \emptyset
    )
    \implies
    \precedes{T}{T'}
    \end{multline*}
\end{proposition}

\begin{preuve}

  \begin{list}{}{}

  \item \textbf{F9.1} If $(T,T')$ is a read-from edge, then \precedes{T}{T'}

    Let $(T,T')$ be a read-from relation.
    By definition it exists a site $i$, a write $w[x] \in T$, and a read $r[x] \in T'$
    such that during \run at site $i$ $w$ write $x$ then $r$ reads the value written by 
    $w$.
    
    Let $t$ and $t'$ be respectively the times at which these two events occured;
    according to \lsdb:

    \begin{list}{}{}
      \item \textbf{F9.1.1} $\tomdeliverSite{o}{i} <_{\run} t <_{\run} t' $
    \end{list}
    
    Then since $T' \in \mvsg{\run}{\precedesRelation}$, $T' \in \submitted{r}$,
    and by assumption \textbf{A1}, it exists $j \in \replicas{x} \inter \correct{r}$
    such that \tomdeliverSite{o'}{j}.

    Now, $\tomdeliverSite{o}{i} \implies \tomdeliverSite{o}{j}$ by the Uniform Agreement,
    and the Total Order properties of Total Order Multicast.
    Consequently using fact \textbf{F9.1.1},

    \begin{displaymath}
      \neg ( \tomdeliverSite{o'}{i} <_{\run} \tomdeliverSite{o}{i} )
      \implies \tomdeliverSite{o}{j} <_{\run} \tomdeliverSite{o'}{j}
    \end{displaymath} 

    concluding our claim.
     
  \item \textbf{F9.2} If $(T,T)$ is a version-order edge, then \precedes{T}{T}

    Let $T_1, T_2, T_3$ be three transactions committed during \run,
    and suppose that it exists a version-order edge 
    $(T_1,T_2) \in \mvsg{\run}{\precedesRelation}$.

    According to the definition of a version order it follows either:

    \begin{enumerate}

    \item it exists $w_1 \in \wo{T_1}, w_2 \in \wo{T_2}$,
      and $r_3 \in \ro{T_3}$  such that $r_3[x_3]$, $w_1[x_1]$ and 
      $x_1 \versionOrder x_2$.
      
      By definition of $x_1 \versionOrder x_2 \implies \precedes{T_1}{T_2}$.

    \item it exists $r_1 \in \ro{T_1}, w_2 \in \wo{T_2}$ and 
      $w_3 \in \wo{T_3}$ such that $r_1[x_3]$, $w_2[x_2]$ and
      $x_3 \versionOrder x_2$.

      Let $i \in \replicas{x} \inter \correct{\run}$ ( by assumption \textbf{A1}).
      Since $T_1, T_2, T_3 \in \committed{r} \subseteq \submitted{r}$,
      $i$ TO-delivers $r_1$, $w_1$ and $w_3$ during \run.
      Now according to the Total Order property of Total Order Multicast,
      since $x_3 \versionOrder x_2$, \precedesSite{w_3}{w_2}{i}.

      Let $j$ be a site on which $r_1[x3]$ happens.
      Since \precedesSite{w_3}{w_2}{i}, according to our definition 
      of \commit{} (Section \ref{sect:commit}), $w_2 \hb r_1$.

      Now since $T_1 \in \committed{\run}$, necessarily \precedesI{r_1}{w_2}
      (otherwise $T_1$ is aborted: line \ref{alg:lsdb:4c}).

    \end{enumerate}

  \end{list}

  By facts \textbf{F9.1} and \textbf{F9.2}

\end{preuve}

\begin{proposition}
  \label{prop:10}
  \run is serializable.
  \bigskip
\end{proposition}

\begin{preuve}

  Consider the sub-graph $G_u$ of \mvsg{r}{\versionOrder} containing all the transactions $T$
  such that $\wo{T} \neq \emptyset$, and the edge linking them.

  \begin{list}{}{}
    
  \item \textbf{F10.1} $G_u$ is acyclic.
    \begin{list}{}{}
    \item Let $T_1,\ldots,T_m \in G_u$ such that $T_1,\ldots,T_{m \geq 1}$ forms a cycle in $G_u$,
      and recall that by definition $T_1,\ldots,T_m \in \committed{r}$

      According to proposition \textbf{P9},
      $T_1 \precedesRelation \ldots \precedesRelation T_m \precedesRelation T_1$.

      Let $i$ be a replica of $T_1$, and we note $t$ the time at which $i$ commits
      $T$ during \run.

      Acoording to \lsdb at time $t$, \closed{T_1}{\Git}.

      Now according to Algorithm~\ref{alg:decide}, and since $i$ commits $T_1$  at time $t$,

      \begin{displaymath}
        \exists k \in \llbracket 2, m \rrbracket, T_k \in \breakCycles{\predecessors{T_1}{\Git}}
      \end{displaymath}

      Let $j \in \replicas{T_k}$ such that $j$ commit $T_k$ during \run, and let $t"$ be
      the time at which this event happens.

      Since $T_1 \in \predecessors{T_k}{\pgraphSiteTime{j}{t'}}$
      and $T_k \in \predecessors{T_1}{\pgraphSiteTime{i}{t}}$,
      \predecessors{T_1}{\Git} = \predecessors{T_k,\pgraphSiteTime{j}{t'}}.

      Consequently since \breakCycles{} is deterministic,
      $j$ cannot commit $T_k$ during \run. Absurd.

    \end{list}

  \item  \textbf{F10.2} $\mvsg{\run}{\precedesRelation}$ is acyclic.
    \begin{list}{}{}
    \item By fact \textbf{F10.1} and since read only transactions are executed using two-phases locking.
    \end{list}
    
  \end{list}

  Fact \textbf{F10.2} induces that \run is serializable.
    
\end{preuve}